\documentclass{ieeeaccess}

\usepackage[table]{xcolor}
\usepackage{graphicx}
\usepackage{textcomp}
\usepackage{algorithm}

\usepackage{algpseudocode}

\usepackage{cite}
\usepackage{amsmath,amssymb,amsfonts}

\newtheorem{definition}{Definition}

\definecolor{matchColorAES}{RGB}{243,176,90}
\definecolor{matchColorWhirl}{RGB}{10,189,60}
\definecolor{matchColorFantomas}{RGB}{2,148,165}
\definecolor{matchColorSkipjack}{RGB}{213,11,83}

\usepackage{tikz}
\usetikzlibrary{matrix, positioning}
 \NewSpotColorSpace{PANTONE}
  \AddSpotColor{PANTONE} {PANTONE3015C} {PANTONE\SpotSpace 3015\SpotSpace C} {1 0.3 0 0.2}
  \SetPageColorSpace{PANTONE}%

\def\BibTeX{{\rm B\kern-.05em{\sc i\kern-.025em b}\kern-.08em
    T\kern-.1667em\lower.7ex\hbox{E}\kern-.125emX}}
\begin{document}
\history{Date of publication xxxx 00, 0000, date of current version xxxx 00, 0000.}
\doi{10.1109/ACCESS.2020.DOI}

\title{On the Design of Chaos-Based S-boxes}
\author{\uppercase{Miroslav M. Dimitrov}\authorrefmark{1}}

\address[1]{Bulgarian Academy of Sciences, Institute of Mathematics and Informatics, Sofia, 1113, Bulgaria (e-mail: mirdim@math.bas.bg)}
\tfootnote{This work has been supported by the Bulgarian National Science Fund under contract number DH 12/8, 15.12.2017.}

\markboth
{Author \headeretal: Preparation of Papers for IEEE TRANSACTIONS and JOURNALS}
{Author \headeretal: Preparation of Papers for IEEE TRANSACTIONS and JOURNALS}

\corresp{Corresponding author: Miroslav M. Dimitrov (e-mail: mirdim@math.bas.bg).}

\begin{abstract}
Substitution boxes (S-boxes) are critical nonlinear elements to achieve cryptanalytic resistance of modern block and stream ciphers. Given their importance, a rich variety of S-box construction strategies exists. In this paper, S-boxes generated by using chaotic functions (CF) are analyzed to measure their actual resistance to linear cryptanalysis. The aforementioned papers emphasize on the average nonlinearity of the S-box coordinates only, ignoring the rest of the S-box components in the process. Thus, the majority of those studies should be re-evaluated. Integrating such S-boxes in a given cryptosystem should be done with a considerable caution. Furthermore, we show that in the context of nonlinearity optimization problem the profit of using chaos structures is negligible. By using two heuristic methods and starting from pseudo-random S-boxes, we repeatedly reached S-boxes, which significantly outperform all previously published CF-based S-boxes, in those cryptographic terms, which the aforementioned papers utilize for comparison. Moreover, we have linked the multi-armed bandit problem to the problem of maximizing an S-box average coordinate nonlinearity value, which further allowed us to reach near-optimal average coordinate nonlinearity values significantly greater than those known in literature.
\end{abstract}


\titlepgskip=-15pt

\maketitle

\section{Introduction}
\label{sec:1}

The cryptographic properties of vector boolean functions, or \textbf{S-boxes}, are thoroughly examined by introducing a rich list of desirable parameters an S-box should have in order to guarantee an acceptable resistance to sophisticated cryptographic attacks such as, for example, linear cryptanalysis \cite{Linear}\cite{ATT3}, differential cryptanalysis \cite{ATT1}, boomerang attack \cite{ATT4} or interpolation attack \cite{Interpolation}. Furthermore, S-boxes are widely used in modern cryptographic algorithms like AES \cite{AES}, Whirlpool \cite{Whirlpool}, Camellia \cite{Camellia} and many others. 

Despite the rich variety of proposed methods for S-boxes generation, we mainly focus on S-box constructions benefiting from the study of chaos, to further analyze their actual resistance to linear cryptanalysis. 

In Section \ref{sec:2} we introduce the definitions of some basic cryptographic characteristics used to measure the cryptographic strength of a given S-box.

In Section \ref{sec:3} we show that the actual nonlinearity value, or \textbf{NL}, of the majority of chaotic functions-based (\textbf{CF}-based) published S-boxes differs from the average nonlinearity value originally announced. This discrepancy is based on the fact that the aforementioned papers consider the average nonlinearity of the S-box coordinates only, or \textbf{ACNV}, ignoring the rest of the S-box components in the process. In Section \ref{sec:4}, we propose an algorithm, which significantly outperforms all previously published S-boxes in terms of ACNV. During our experiments, we repeatedly reached S-boxes with ACNV of 114. We want to emphasize, that ACNV greater than 112.0, to the best of our knowledge, was never achieved in the literature.

In Section \ref{sec:5}, we demonstrate the efficiency of the proposed algorithm by optimizing the ACNV of some popular S-boxes. Thus, we show that the starting state of the optimization routines is negligible. Having this in mind, the competitiveness of S-boxes generated by exploiting chaos structures, at least in the context of S-box nonlinearity optimization problem, is arguable. The same observation was made in \cite{ozkaynak2020effect}. 

Then, in Section \ref{sec:bandits}, we translate the  S-box ACNV optimization problem to the multi-armed bandit problem, which allow us to further improve our results by reaching an ACNV of 114.5 - a value significantly larger than those known in literature.


\section{Preliminaries}
\label{sec:2}
Let $B = \{0,1\}$. A \textbf{Boolean function} $f(x)$ of $n$ variables $x_1, \cdots,x_n$ is a mapping $f : B^n \mapsto B$ from $n$ binary inputs $x=(x_1,x_2,\cdots,x_n) \in B^n$ to one binary output $y = f(x) \in B$.

\begin{definition}[Algebraic Normal Form -- ${ANF}_f$]
The algebraic normal form of an $n$-variable Boolean function $f(x)$ is given by the following equation: \[ {ANF}_f = a_0 \oplus a_1x_1 \oplus a_2x_2 \oplus a_{1,2,\cdots,n}x_1x_2 \cdots x_n, \] where the coefficients $a_{i \cdots j} \in B = \{0,1\}$.
\end{definition}

A linear Boolean function $f$ is a function with specific algebraic normal form $ANF_f$, s.t. no term with algebraic degree greater than 1 exists. A more formal definition follows: 

\begin{definition}[Linear Boolean Function]
Any $n$-variable Boolean function of the form: $$l_w(x) = <w,x> = w_1x_1 \oplus w_2x_2 \oplus \cdots \oplus w_{n}x_{n},$$ where $w, x \in B^n$, is called a linear Boolean function.
\end{definition}

An $n$-binary input into $m$-binary output mapping $S : B^n \Leftrightarrow B^m$, which assigns some $y = (y_1, y_2, \cdots , y_m) \in B^m$ by $S(x) = y$ to each $x = (x_1, x_2, \cdots, x_n) \in B^n$, is called an ($n \times m$) substitution table (\textbf{S-box}) and is denoted by $S(n, m)$.

An S-box $S(n, m)$ is said to be \textbf{bijective}, if it maps each input $x \in B^n$ to a distinct output $y = S(x) \in B^m$ and all possible $2^m$ outputs are present. For example, the $(n,n)$ bijective S-boxes are Boolean permutations on $F_2^n$, where $F_2$ is a finite field with two elements.

An S-box $S(n, m)$ can be bijective only when $n = m$. Clearly, a bijective S-box $S(n, n)$ represents a permutation of its $2^n$ inputs, since each input is mapped to a distinct output and all possible $2^n$ outputs are present. In this way, $S(n, n)$ will be reversible, that is, there is a mapping from each distinct output to its corresponding input.

\begin{definition}[Look-up Table]
The \textbf{look-up table LUT} of an S-box $S(n, m)$ is a ($2^n \times m$) binary matrix $S_{LUT}$, which rows consist of all outputs of $S(n,m)$, corresponding to all possible $2^n$ inputs ordered lexicographically.
\[ S = \begin{bmatrix}
f_1(0,0,...,0) & f_2(0,0, ...,0) & ... &  & f_m(0,0,...,0)\\
f_1(0,0,...,1) & f_2(0,0,...,1) & \cdots &  & f_m(0,0,...,1) \\
 \vdots & \vdots & \vdots &  & \vdots \\
f_1(1,1, ... ,0) & f_2(1,1, ... ,0) & \cdots &  & f_m(1,1, ..., 0) \\
f_1(1,1,...,1) & f_2(1,1,...,1) & \cdots &  & f_m(1,1,...,1)
\end{bmatrix} \]

We define each column of $S_{LUT}$ as \textbf{coordinate} of S. All linear combinations of coordinates of S are called \textbf{components} of S. 
\end{definition}

\begin{definition}[Linear Approximation Table]
The linear approximation table of an S-box $S(n, m)$, denoted by $S_{LAT}$, is a $(2^n \times 2^m)$ table, which entries are given by: ${S}_{LAT}[X][Y] = 2^{n-1}-d_H(X,Y)$, where $Y$ is a linear combination of the coordinates of the current S-box, $X$ is the consequent linear function with length $n$ and $d_H(X,Y)$ denotes the Hamming distance between $X$ and $Y$.
\end{definition}

The linear approximation table of a given S-box S(n,n) reveals the actual correlation between the components of S and all linear Boolean functions sharing the same dimension n.   

\begin{definition}[S-box Nonlinearity]
\label{def:S-boxNonlinearity}
The nonlinearity of an S-box $S(n, m)$ is defined as $S_{NL} = 2^{n-1} - max\left( \{abs(w_i)  \}\right),$ where $\left \{ w_i \right \}$ is the set of all elements in the LAT, excluding the first row and the first column.
\end{definition}

Lower values of nonlinearity could be exploited by the family of linear cryptanalysis attacks. Having this in mind, higher nonlinearity value is a desirable S-box property.

Each S-box is uniquely defined by its LUT. Therefore, if we translate each row of the LUT as decimal number, we can obtain a unique decimal representation of the S-box denoted by \textbf{DLUT}.



\section{Chaos-based S-box constructions}
\label{sec:3}

The methods involved in CF S-box constructions are manifold. For example, chaos function combined with travelling salesman problem \cite{ahmad2016efficient}, chaotic substitution box design \cite{ahmad2014new}, 1D chaotic map combined with $\beta$-Hill climbing \cite{alzaidi2018new}, chaotic map combined with sine-cosine optimization \cite{alzaidi2018sine}, chaotic system with multiple attractors \cite{lai2018new}, chaotic map combined with heuristics \cite{belazi2017efficient}, one-dimensional discrete chaotic map \cite{lambic2018s}, hyperchaotic systems \cite{peng2016novel} \cite{al2018new}, spatiotemporal chaotic dynamics \cite{liu2018novel}, chaotic map combined with genetic algorithms \cite{wang2012novel}, chaotic logistic maps combined with bacterial foraging optimization \cite{tian2017chaotic} and many others (see Table \ref{tab:summaryResults}). Usually, the best candidate of each method is further compared to others in terms of important cryptographic properties like nonlinearity, differential uniformity \cite{carlet2010vectorial} and strict avalanche criterion (SAC) \cite{forrie1988strict}. The majority of authors emphasize on the ACNV of their best candidate. In Table \ref{tab:comparison8} the coordinate nonlinearities of several S-box candidates achieved by some CF-based methods are presented. A more detailed overview is given in Table \ref{tab:summaryResults}.

\begin{figure*}
\centering
  \includegraphics[width=\textwidth]{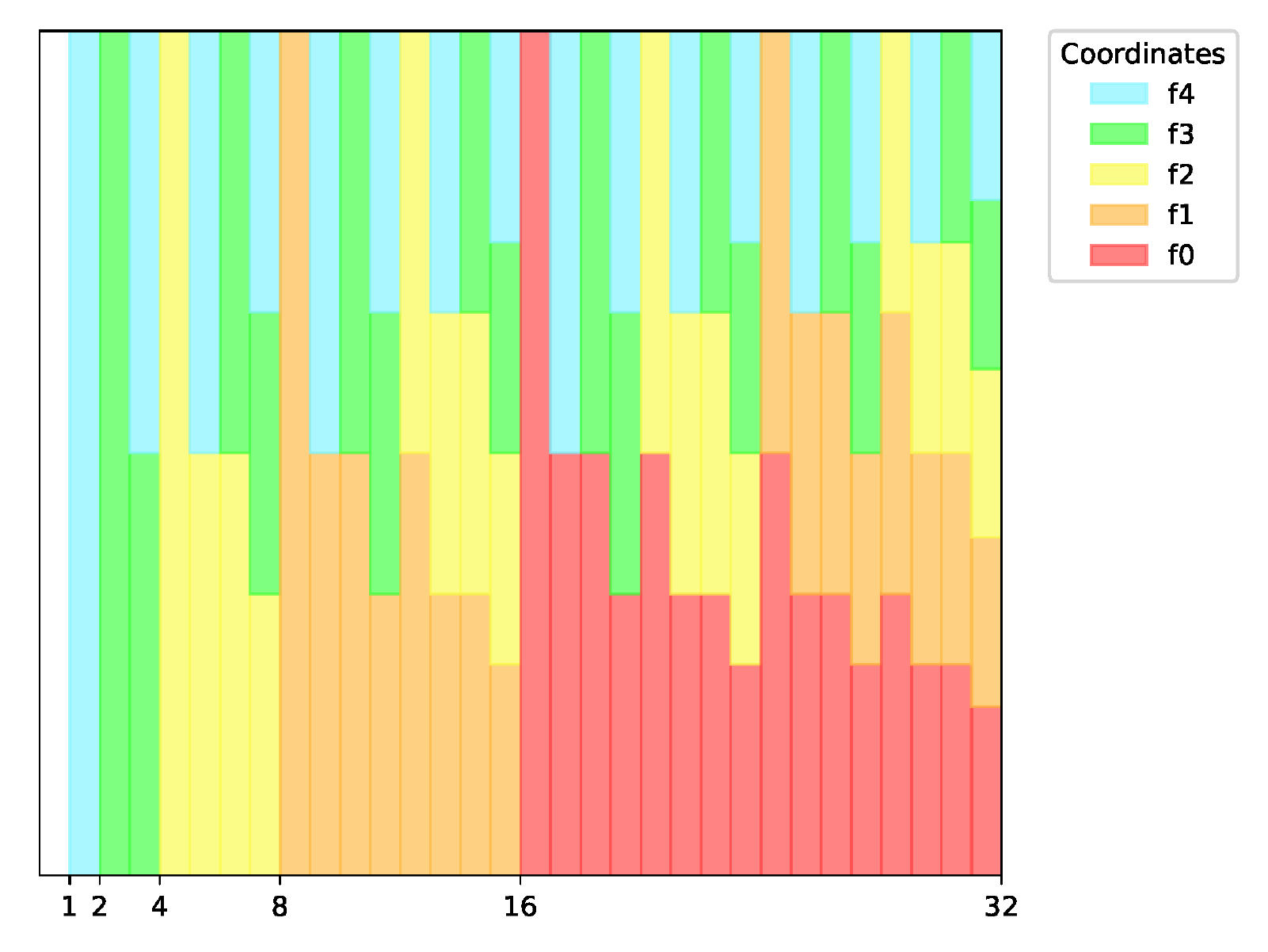}
\caption{Coordinate decomposition of a $(5,5)$ S-box LAT}
\label{fig:1}       
\end{figure*}

The actual nonlinearity of an S-box is calculated by the minimum nonlinearity of all the components of the S-box. For example, let us take an arbitrary S-box $F(5,5)$ with $F_{LUT} = [f0,f1,f2,f3,f4]$. Each column of $F_{LAT}$ is determined by some linear combination of coordinates of $F$, sorted lexicographically, from left to right, by the binary representation of the column index, zero-filled to 5. Let $F_{LAT}[i]$ denotes the $i$-th column of $F_{LAT}$. Then, for example, the $F_{LAT}[11]$ column holds the nonlinear characteristics of the Boolean function $f_1 \oplus f_3 \oplus f_4$, while $F_{LAT}[4]$ holds the nonlinear characteristics of the Boolean function $f_3$. In Figure \ref{fig:1} the coordinate decomposition of $F_{LAT}$ is visualized. Each coordinate is associated with distinct color. The number of segments in each column corresponds to the number of terms in the respective linear combination of coordinates. Since $F_{LAT}[0]$ is the trivial linear combination (all coefficients are equal to zero), we leave the first column of Figure \ref{fig:1} colorless. For technical reasons and better illustration, the coordinate decomposition example is based on a $(5,5)$ S-box. However, it is applicable to S-boxes of any dimension.

As defined in Definition \ref{def:S-boxNonlinearity}, we seek the maximum absolute value $v$ of all the elements in S-box $S(n,n)$ LAT, to find the nonlinearity of $S$, i.e. $S_{NL} = 2^{n-1} - v$.

In Table \ref{tab:comparisonNL} the actual nonlinearity of each S-box from Table \ref{tab:comparison8} is calculated. The deviations observed are due to the fact that the designers consider the nonlinearity values of coordinates only (the non-segmented columns in the $(8,8)$ coordinate decomposition).

In the context of block ciphers, a low nonlinearity S-box value is associated with the cipher linear cryptanalysis resistance \cite{matsui1993linear}\cite{ATT3}\cite{heys2002tutorial}. 

\begin{table}
\centering
\caption{Comparison of nonlinearity of some CF-based $(8,8)$ S-boxes}
\label{tab:comparison8}       
\scalebox{0.9}{
\begin{tabular}{rllllllll}
 & \multicolumn{8}{c}{} \\
 Method & $f_1$ & $f_2$ & $f_3$ & $f_4$ & $f_5$ & $f_6$ & $f_7$ & $f_8$ \\
\noalign{\smallskip}\hline\noalign{\smallskip}
 Ahmad \cite{ahmad2016efficient} & 108 & 110 & 110 & 108 & 106 & 106 & 106 & 106 \\
 Ahmad \cite{ahmad2014new} & 104 & 106 & 106 & 104 & 102 & 108 & 106 & 106 \\
 Al Solami \cite{al2018new} & 108 & 110 & 108 & 108 & 106 & 110 & 108 & 110 \\
 Alzaidi \cite{alzaidi2018new} & 110 & 112 & 110 & 110 & 110 & 110 & 110 & 110 \\
 Alzaidi \cite{alzaidi2018sine} & 110 & 110 & 110 & 110 & 110 & 108 & 110 & 108 \\
 Belazi \cite{belazi2017efficient} & 106 & 106 & 106 & 104 & 108 & 102 & 106 & 104 \\
 Lai \cite{lai2018new} & 104 & 110 & 104 & 108 & 104 & 104 & 106 & 104 \\
 Liu \cite{liu2018novel} & 108 & 102 & 104 & 104 & 102 & 104 & 106 & 106 \\
 Lambic \cite{lambic2018s} & 106 & 106 & 106 & 106 & 106 & 108 & 108 & 106 \\
 Peng \cite{peng2016novel} & 102 & 102 & 104 & 104 & 102 & 100 & 106 & 102 \\
 Tian \cite{tian2017chaotic} & 106 & 106 & 110 & 108 & 106 & 108 & 108 & 108 \\

\noalign{\smallskip}\hline
\end{tabular}}
\end{table}

\begin{table}
\begin{center}
\caption{Real nonlinearity values (NL) of the S-boxes given in Table \ref{tab:comparison8}}
\label{tab:comparisonNL}       
\scalebox{1}{
\begin{tabular}{rllrc}
\hline
 Method & min & max & ACNV & NL \\
\noalign{\smallskip}\hline\noalign{\smallskip}
 Ahmad \cite{ahmad2016efficient} & 106 & 110 & 107.5 & 90 \\
 Ahmad \cite{ahmad2014new} & 102 & 108 & 105.25 & 94\\
 Al Solami \cite{al2018new} & 106 & 110 & 108.5 & 94 \\
 Alzaidi \cite{alzaidi2018new} & 110 & 112 & 110.25 & 96 \\
 Alzaidi \cite{alzaidi2018sine} & 108 & 110 & 109.5 & 94 \\
 Belazi \cite{belazi2017efficient} & 102 & 108 & 105.25 & 88 \\
 Lai \cite{lai2018new} & 104 & 110 & 105.5 & 92 \\
 Liu \cite{liu2018novel} & 102 & 108 & 104.5 & 96 \\
 Lambic \cite{lambic2018s} & 106 & 108 & 106.5 & 94 \\
 Peng \cite{peng2016novel} & 100 & 102 & 102.75 & 88 \\
 Tian \cite{tian2017chaotic} & 106 & 110 & 107.5 & 92 \\
\noalign{\smallskip}\hline
\end{tabular}}
\end{center}
\end{table}

\section{Alternative construction}
\label{sec:4}

As we have shown in the previous section, the average value of the nonlinearities of the coordinates of a given S-box $S$ doesn't correspond to the the actual nonlinearity of $S$. However, from the designer perspective, if a higher value of ACNV is desirable, a new  heuristic construction is suggested. 

\begin{figure}
\centering
  \includegraphics[width=0.5\textwidth]{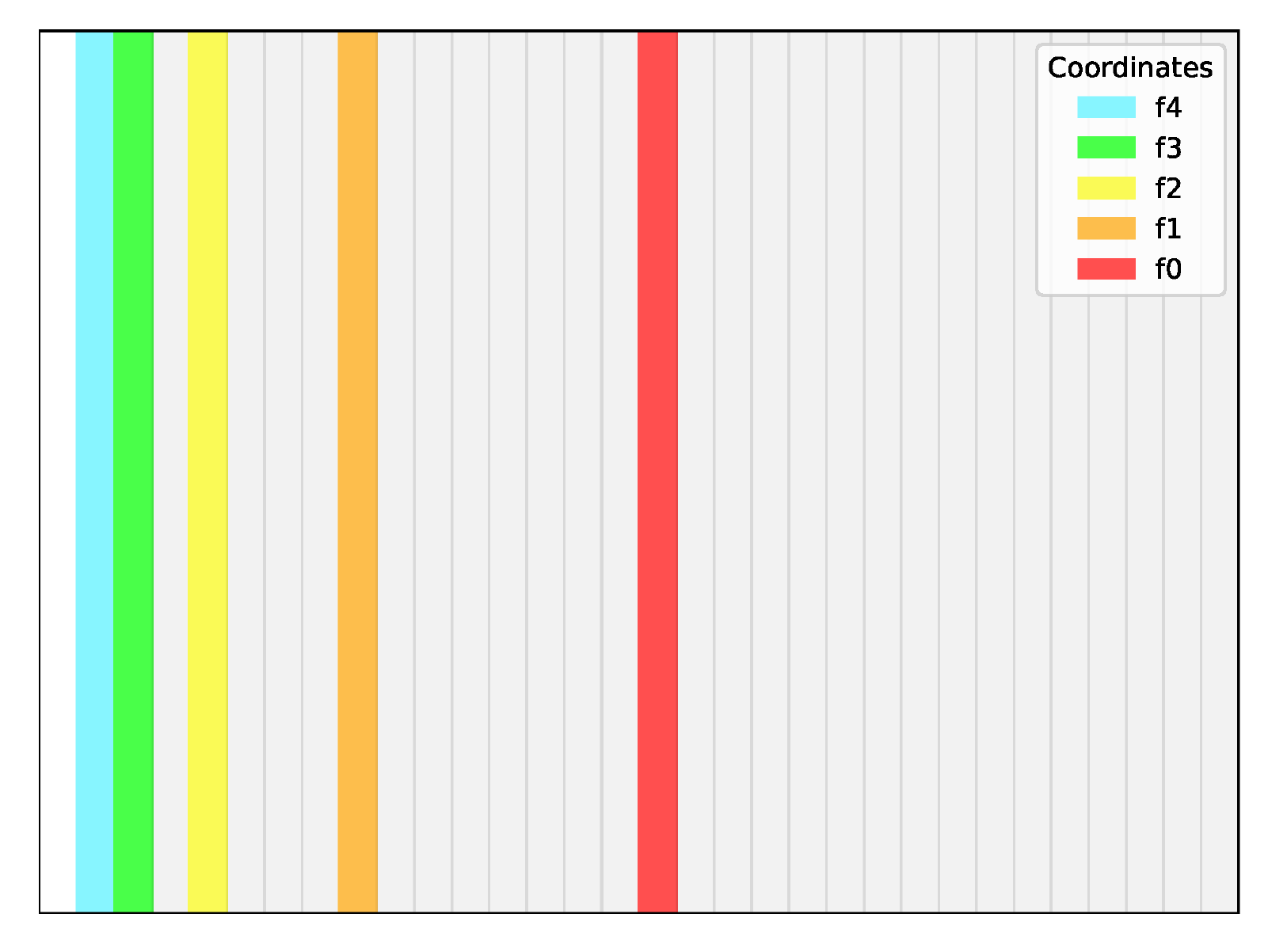}
\caption{Columns of interest of a $(5,5)$ S-box LAT}
\label{fig:2}       
\end{figure}

In general, if we want to improve the nonlinearity of a given  bijective S-box $S(n,n)$, a strategy of lowering the absolute value of coefficients in $S_{LAT}$ makes sense. Moreover, the elements of each column of $S_{LAT}$ are entangled by the Parceval's theorem \cite{meier1989nonlinearity}. Let's denote as $C_i$ the array composed of the elements of $S_{LAT}[i]$. Since we want to lower the nonlinearities of coordinates of $S$ only, an evaluating function $E(S)$ is created, s.t. $ E(S) = \sum_{p=0}^{n-1}\sum_{x \in C_{2^p}} { \lvert x \rvert ^ {M}},$ where $M$ denotes a magnitude of our choice. The restriction $x \in C_{2^p}$ narrows down the set of possible columns of $S_{LAT}$ to be optimized, in terms of nonlinearity, to the set of coordinates of $S$. As example, in case of a $S(5,5)$ S-box, the evaluation function threats as significant the elements inside the colored columns of $S_{LAT}$ illustrated in Figure \ref{fig:2}.

By using stochastic\footnote{hill climbing without neighborhood search} hill climbing as heuristic function, starting from arbitrary pseudo-random S-box construction and by using $E(S)$, algorithm \ref{algo} is proposed.

\begin{algorithm}[H]
\caption{}\label{algo}
\begin{algorithmic}[1]
\State $s\gets R(n)$ \Comment{the function R(n) generates pseudo-random bijective S-box $S(n,n)$}
\Repeat
	\State $sdupl\gets s$
	\State \Call{RT}{sdupl} \Comment{the function RT(S) make a random transposition in $S$}
	
	\If{$\Call{E}{sdupl}  < \Call{E}{s}$}
		\State $s\gets sdupl$
	\EndIf	
\Until{STOP condition is reached}  \Comment{reaching $\frac{n(n-1)}{4}$ cycles}

\end{algorithmic}
\end{algorithm}

Given an S-box $S(n,n)$, and by using just one transposition, we can reach a total of ${n\choose 2}$ S-boxes. Let denote this set as $S^T$. We further define a set $S^I$, s.t. $W \in S^I \iff W \in S^T \land E(W) < E(S)$. In case $\lvert S^I \rvert = 1$, and we are allowed to randomly pick $ \frac{\lvert S^T \rvert} {2}$ elements from $S^T$, the probability some of the picked elements to belong to $S^I$ is $\frac{1}{2}$. The threshold value of the stop condition in Algorithm \ref{algo} is constructed on this observation.

\section{Results Part I}
\label{sec:5}
By using a magnitude of $10$, we repeatedly generated S-boxes with high coordinate nonlinearities. During our experiments, we have tried various magnitude values. However, larger or smaller values of the magnitude are respectively too aggressive or too tolerant to the largest elements of the S-box LAT.

In Figure \ref{fig:Sc_raw} the DLUT, in a hexadecimal format, of an optimized S-box $S_c(8,8)$ is presented. The first row and column of the table correspond respectively to the first and second half of the input in hexadecimal format. For example, the input \textbf{11110101}, equal to \textbf{f5}, is transformed by $S_c$ to \textbf{5d}. The characteristics of $S_c$ are summarized in Tables \ref{tab:proposed8} and \ref{tab:proposedNL}.

In \cite{tanyildizi2019new}, Table 5, a summary on the CF-based S-box constructions found in the literature is presented (an updated version of it is to be found in Table \ref{tab:summaryResults}). We significantly outperform all of them in terms of ACNV and SAC, reaching the optimal SAC value of 0.5.

We further launched the algorithm on some popular (8,8) S-box constructions. However, because of the non deterministic nature of the optimization process, it is difficult to match a given S-box input $S_{start}$, which is to be optimized, with the final optimized S-box $S_{end}$. To achieve such matching, we have restricted the algorithm of changing the first 16 elements of $S_{start}$. This allows us to further demonstrate the flexibility of the optimization process. Furthermore, since the first 16 elements of $S_{start}$ and $S_{end}$ are always shared, $S_{end}$ can be successfully matched to $S_{start}$. 

In Figures \ref{fig:S_AES_opt} and \ref{fig:S_Whirl_opt}, an optimized by algorithm \ref{algo} versions of Rijndael \cite{AES} and Whirlpool \cite{Whirlpool} S-boxes are presented. The colored cells represent those elements of the corresponding S-box, which were not modified during the optimization process. Furthermore, in Figures \ref{fig:S_Fantomas_opt} and \ref{fig:S_Skipjack_opt}, the optimized versions of Fantomas \cite{grosso2014ls} and Skipjack \cite{Skipjack} S-boxes are given. 

 All of the aforementioned S-boxes are optimized to the ACNV of 114.0. Algorithm \ref{algo} was implemented with the built-in tools provided by the open-source mathematical software system SageMath \cite{developers2016sagemath}.

\begin{figure}
\centering

\begin{tikzpicture}[scale=0.7, every node/.style={anchor=base,text depth=.4ex,text height=1ex,text width=2ex, scale=0.7}]

\matrix (A) [matrix of nodes, nodes={draw}, font=\ttfamily, row 1/.style={nodes={fill=gray!50}}, column 1/.style={nodes={fill=gray!50}}]
	{	& 00 & 01 & 02 & 03 & 04 & 05 & 06 & 07 & 08 & 09 & 0a & 0b  & 0c & 0d & 0e & 0f \\
		00 & |[fill=matchColorAES!50]| 63 & |[fill=matchColorAES!50]| 7c & |[fill=matchColorAES!50]| 77 & |[fill=matchColorAES!50]| 7b & |[fill=matchColorAES!50]| f2 & |[fill=matchColorAES!50]| 6b & |[fill=matchColorAES!50]| 6f & |[fill=matchColorAES!50]| c5 & |[fill=matchColorAES!50]| 30 & |[fill=matchColorAES!50]| 01 & |[fill=matchColorAES!50]| 67 & |[fill=matchColorAES!50]| 2b & |[fill=matchColorAES!50]| fe & |[fill=matchColorAES!50]| d7 & |[fill=matchColorAES!50]| ab & |[fill=matchColorAES!50]| 76 \\
10 & c2 & 92 & |[fill=matchColorAES!50]| c9 & |[fill=matchColorAES!50]| 7d & fb & |[fill=matchColorAES!50]| 59 & |[fill=matchColorAES!50]| 47 & |[fill=matchColorAES!50]| f0 & |[fill=matchColorAES!50]| ad & c4 & c0 & ae & c8 & |[fill=matchColorAES!50]| a4 & 3a & a8 \\
20 & |[fill=matchColorAES!50]| b7 & |[fill=matchColorAES!50]| fd & |[fill=matchColorAES!50]| 93 & |[fill=matchColorAES!50]| 26 & |[fill=matchColorAES!50]| 36 & |[fill=matchColorAES!50]| 3f & fa & |[fill=matchColorAES!50]| cc & |[fill=matchColorAES!50]| 34 & |[fill=matchColorAES!50]| a5 & |[fill=matchColorAES!50]| e5 & |[fill=matchColorAES!50]| f1 & |[fill=matchColorAES!50]| 71 & |[fill=matchColorAES!50]| d8 & b5 & 11 \\
30 & |[fill=matchColorAES!50]| 04 & |[fill=matchColorAES!50]| c7 & 33 & |[fill=matchColorAES!50]| c3 & |[fill=matchColorAES!50]| 18 & |[fill=matchColorAES!50]| 96 & |[fill=matchColorAES!50]| 05 & 9c & |[fill=matchColorAES!50]| 07 & |[fill=matchColorAES!50]| 12 & 88 & e1 & |[fill=matchColorAES!50]| eb & 2f & |[fill=matchColorAES!50]| b2 & 35 \\
40 & |[fill=matchColorAES!50]| 09 & 82 & |[fill=matchColorAES!50]| 2c & 1b & 19 & |[fill=matchColorAES!50]| 6e & 1a & |[fill=matchColorAES!50]| a0 & 42 & |[fill=matchColorAES!50]| 3b & |[fill=matchColorAES!50]| d6 & bb & |[fill=matchColorAES!50]| 29 & e2 & 27 & 85 \\
50 & |[fill=matchColorAES!50]| 53 & |[fill=matchColorAES!50]| d1 & 20 & |[fill=matchColorAES!50]| ed & 10 & |[fill=matchColorAES!50]| fc & 31 & 4b & |[fill=matchColorAES!50]| 6a & |[fill=matchColorAES!50]| cb & |[fill=matchColorAES!50]| be & |[fill=matchColorAES!50]| 39 & 8e & |[fill=matchColorAES!50]| 4c & |[fill=matchColorAES!50]| 58 & ef \\
60 & |[fill=matchColorAES!50]| d0 & 8f & a2 & f7 & |[fill=matchColorAES!50]| 43 & 0d & 23 & 84 & 46 & e9 & 03 & 7e & |[fill=matchColorAES!50]| 50 & 38 & |[fill=matchColorAES!50]| 9f & 2a \\
70 & |[fill=matchColorAES!50]| 51 & |[fill=matchColorAES!50]| a3 & |[fill=matchColorAES!50]| 40 & cf & 83 & |[fill=matchColorAES!50]| 9d & 28 & f3 & bd & |[fill=matchColorAES!50]| b6 & de & |[fill=matchColorAES!50]| 21 & 00 & |[fill=matchColorAES!50]| ff & f5 & |[fill=matchColorAES!50]| d2 \\
80 & |[fill=matchColorAES!50]| cd & d4 & |[fill=matchColorAES!50]| 13 & ee & 5d & |[fill=matchColorAES!50]| 97 & |[fill=matchColorAES!50]| 44 & |[fill=matchColorAES!50]| 17 & 0c & |[fill=matchColorAES!50]| a7 & 7f & |[fill=matchColorAES!50]| 3d & |[fill=matchColorAES!50]| 64 & 5f & 5b & |[fill=matchColorAES!50]| 73 \\
90 & 41 & a1 & |[fill=matchColorAES!50]| 4f & dd & |[fill=matchColorAES!50]| 22 & aa & |[fill=matchColorAES!50]| 90 & 80 & 45 & ec & f8 & |[fill=matchColorAES!50]| 14 & db & |[fill=matchColorAES!50]| 5e & |[fill=matchColorAES!50]| 0b & 9b \\
a0 & |[fill=matchColorAES!50]| e0 & 7a & 6d & |[fill=matchColorAES!50]| 0a & |[fill=matchColorAES!50]| 49 & |[fill=matchColorAES!50]| 06 & |[fill=matchColorAES!50]| 24 & |[fill=matchColorAES!50]| 5c & ca & |[fill=matchColorAES!50]| d3 & |[fill=matchColorAES!50]| ac & 66 & |[fill=matchColorAES!50]| 91 & |[fill=matchColorAES!50]| 95 & |[fill=matchColorAES!50]| e4 & 32 \\
b0 & |[fill=matchColorAES!50]| e7 & 9a & |[fill=matchColorAES!50]| 37 & 79 & |[fill=matchColorAES!50]| 8d & |[fill=matchColorAES!50]| d5 & |[fill=matchColorAES!50]| 4e & |[fill=matchColorAES!50]| a9 & |[fill=matchColorAES!50]| 6c & |[fill=matchColorAES!50]| 56 & |[fill=matchColorAES!50]| f4 & |[fill=matchColorAES!50]| ea & |[fill=matchColorAES!50]| 65 & 72 & af & |[fill=matchColorAES!50]| 08 \\
c0 & |[fill=matchColorAES!50]| ba & |[fill=matchColorAES!50]| 78 & |[fill=matchColorAES!50]| 25 & |[fill=matchColorAES!50]| 2e & |[fill=matchColorAES!50]| 1c & |[fill=matchColorAES!50]| a6 & |[fill=matchColorAES!50]| b4 & |[fill=matchColorAES!50]| c6 & |[fill=matchColorAES!50]| e8 & dc & |[fill=matchColorAES!50]| 74 & |[fill=matchColorAES!50]| 1f & 5a & bc & |[fill=matchColorAES!50]| 8b & |[fill=matchColorAES!50]| 8a \\
d0 & |[fill=matchColorAES!50]| 70 & |[fill=matchColorAES!50]| 3e & b1 & e6 & |[fill=matchColorAES!50]| 48 & 02 & |[fill=matchColorAES!50]| f6 & |[fill=matchColorAES!50]| 0e & |[fill=matchColorAES!50]| 61 & 75 & |[fill=matchColorAES!50]| 57 & |[fill=matchColorAES!50]| b9 & |[fill=matchColorAES!50]| 86 & |[fill=matchColorAES!50]| c1 & |[fill=matchColorAES!50]| 1d & |[fill=matchColorAES!50]| 9e \\
e0 & e3 & b8 & |[fill=matchColorAES!50]| 98 & 15 & |[fill=matchColorAES!50]| 69 & |[fill=matchColorAES!50]| d9 & 4a & |[fill=matchColorAES!50]| 94 & da & |[fill=matchColorAES!50]| 1e & |[fill=matchColorAES!50]| 87 & f9 & |[fill=matchColorAES!50]| ce & |[fill=matchColorAES!50]| 55 & 3c & |[fill=matchColorAES!50]| df \\
f0 & |[fill=matchColorAES!50]| 8c & 81 & |[fill=matchColorAES!50]| 89 & 4d & |[fill=matchColorAES!50]| bf & 62 & 52 & |[fill=matchColorAES!50]| 68 & 60 & |[fill=matchColorAES!50]| 99 & |[fill=matchColorAES!50]| 2d & |[fill=matchColorAES!50]| 0f & |[fill=matchColorAES!50]| b0 & |[fill=matchColorAES!50]| 54 & b3 & |[fill=matchColorAES!50]| 16 \\
 	};
\end{tikzpicture}
\caption{An optimized AES S-box using Algorithm \ref{algo}}
\label{fig:S_AES_opt}
\end{figure}

\begin{figure}
\centering

\begin{tikzpicture}[scale=0.7, every node/.style={anchor=base,text depth=.4ex,text height=1ex,text width=2ex, scale=0.7}]

\matrix (A) [matrix of nodes, nodes={draw}, font=\ttfamily, row 1/.style={nodes={fill=gray!50}}, column 1/.style={nodes={fill=gray!50}}]
	{	& 00 & 01 & 02 & 03 & 04 & 05 & 06 & 07 & 08 & 09 & 0a & 0b  & 0c & 0d & 0e & 0f \\
		00 & |[fill=matchColorWhirl!50]| 18 & |[fill=matchColorWhirl!50]| 23 & |[fill=matchColorWhirl!50]| c6 & |[fill=matchColorWhirl!50]| e8 & |[fill=matchColorWhirl!50]| 87 & |[fill=matchColorWhirl!50]| b8 & |[fill=matchColorWhirl!50]| 01 & |[fill=matchColorWhirl!50]| 4f & |[fill=matchColorWhirl!50]| 36 & |[fill=matchColorWhirl!50]| a6 & |[fill=matchColorWhirl!50]| d2 & |[fill=matchColorWhirl!50]| f5 & |[fill=matchColorWhirl!50]| 79 & |[fill=matchColorWhirl!50]| 6f & |[fill=matchColorWhirl!50]| 91 & |[fill=matchColorWhirl!50]| 52 \\
10 & f8 & 46 & ae & 1f & df & 5d & 0d & 83 & ca & 28 & e0 & ec & 22 & 6d & |[fill=matchColorWhirl!50]| fe & 2a \\
20 & 53 & 67 & |[fill=matchColorWhirl!50]| 37 & f4 & 3f & d0 & 2d & ce & 38 & 59 & 2b & 0c & c9 & e3 & |[fill=matchColorWhirl!50]| 6b & b1 \\
30 & 9d & 5b & |[fill=matchColorWhirl!50]| 10 & 26 & |[fill=matchColorWhirl!50]| cb & bc & da & 73 & e6 & 33 & |[fill=matchColorWhirl!50]| 41 & 09 & 34 & |[fill=matchColorWhirl!50]| 7d & fc & 98 \\
40 & 92 & dd & 74 & 35 & bb & ab & 5f & |[fill=matchColorWhirl!50]| 9e & c8 & 5c & a7 & 8d & af & 4c & 81 & 3b \\
50 & 19 & 32 & 0b & |[fill=matchColorWhirl!50]| 71 & 8f & 89 & 13 & 63 & c2 & 51 & 4a & 21 & |[fill=matchColorWhirl!50]| 9a & 0a & 06 & 04 \\
60 & 93 & |[fill=matchColorWhirl!50]| 0f & |[fill=matchColorWhirl!50]| d5 & 82 & ba & |[fill=matchColorWhirl!50]| cd & 1e & 84 & |[fill=matchColorWhirl!50]| ff & |[fill=matchColorWhirl!50]| 7a & 85 & 47 & e1 & c1 & |[fill=matchColorWhirl!50]| 1a & e4 \\
70 & 6c & |[fill=matchColorWhirl!50]| 54 & c3 & f3 & 60 & ad & 8c & 5e & 1c & 44 & 02 & dc & 1b & |[fill=matchColorWhirl!50]| a1 & a3 & |[fill=matchColorWhirl!50]| 3d \\
80 & b5 & |[fill=matchColorWhirl!50]| 00 & ed & 2f & 78 & d9 & 9f & 49 & 17 & 69 & |[fill=matchColorWhirl!50]| 6a & f7 & 31 & 77 & |[fill=matchColorWhirl!50]| 30 & |[fill=matchColorWhirl!50]| ef \\
90 & 2e & d1 & |[fill=matchColorWhirl!50]| a2 & |[fill=matchColorWhirl!50]| ea & fb & 9b & 97 & 40 & d8 & 9c & 55 & |[fill=matchColorWhirl!50]| 4d & f2 & |[fill=matchColorWhirl!50]| 75 & 27 & |[fill=matchColorWhirl!50]| 8a \\
a0 & eb & 86 & 5a & fa & 42 & ee & |[fill=matchColorWhirl!50]| a8 & b7 & f1 & 57 & 29 & 80 & 90 & 12 & |[fill=matchColorWhirl!50]| 39 & 65 \\
b0 & cf & 3c & 76 & f0 & 7b & 61 & 62 & 03 & 11 & 45 & 20 & 16 & |[fill=matchColorWhirl!50]| 43 & |[fill=matchColorWhirl!50]| c7 & bf & d6 \\
c0 & 05 & 4e & 4b & 1d & db & b3 & |[fill=matchColorWhirl!50]| 7e & f6 & 72 & 66 & de & a5 & e7 & f9 & 7f & b2 \\
d0 & 25 & b0 & be & 50 & 68 & 70 & bd & 07 & d3 & |[fill=matchColorWhirl!50]| 6e & |[fill=matchColorWhirl!50]| c4 & e5 & 3e & b9 & 8e & |[fill=matchColorWhirl!50]| a9 \\
e0 & 8b & 94 & 08 & 15 & cc & aa & fd & b4 & c5 & 58 & a0 & 56 & 2c & 0e & |[fill=matchColorWhirl!50]| 14 & c0 \\
f0 & 24 & |[fill=matchColorWhirl!50]| 3a & ac & 99 & e9 & |[fill=matchColorWhirl!50]| b6 & d4 & 95 & 48 & a4 & 88 & e2 & d7 & 7c & 64 & 96 \\
 	};
\end{tikzpicture}
\caption{An optimized Whirlpool S-box using Algorithm \ref{algo}}
\label{fig:S_Whirl_opt}
\end{figure}

\begin{figure}
\centering

\begin{tikzpicture}[scale=0.7, every node/.style={anchor=base,text depth=.4ex,text height=1ex,text width=2ex, scale=0.7}]

\matrix (A) [matrix of nodes, nodes={draw}, font=\ttfamily, row 1/.style={nodes={fill=gray!50}}, column 1/.style={nodes={fill=gray!50}}]
	{	& 00 & 01 & 02 & 03 & 04 & 05 & 06 & 07 & 08 & 09 & 0a & 0b  & 0c & 0d & 0e & 0f \\
		00 & |[fill=matchColorFantomas!50]| 1e & |[fill=matchColorFantomas!50]| 75 & |[fill=matchColorFantomas!50]| 5f & |[fill=matchColorFantomas!50]| e1 & |[fill=matchColorFantomas!50]| 99 & |[fill=matchColorFantomas!50]| fc & |[fill=matchColorFantomas!50]| 89 & |[fill=matchColorFantomas!50]| 2f & |[fill=matchColorFantomas!50]| 86 & |[fill=matchColorFantomas!50]| ee & |[fill=matchColorFantomas!50]| f1 & |[fill=matchColorFantomas!50]| 7b & |[fill=matchColorFantomas!50]| 23 & |[fill=matchColorFantomas!50]| 52 & |[fill=matchColorFantomas!50]| 10 & |[fill=matchColorFantomas!50]| 94 \\
10 & 4f & 59 & 2c & 8b & f8 & |[fill=matchColorFantomas!50]| 42 & 30 & 00 & 6e & 84 & 35 & 70 & a0 & c3 & 34 & 6f \\
20 & 4e & 41 & 01 & 78 & 8f & |[fill=matchColorFantomas!50]| a8 & 07 & |[fill=matchColorFantomas!50]| 6c & 62 & af & 7f & |[fill=matchColorFantomas!50]| 22 & |[fill=matchColorFantomas!50]| 60 & 79 & 90 & ec \\
30 & 68 & |[fill=matchColorFantomas!50]| f4 & c4 & |[fill=matchColorFantomas!50]| 32 & 1d & 8c & 0e & ce & de & |[fill=matchColorFantomas!50]| 3f & 44 & 1f & 40 & 98 & |[fill=matchColorFantomas!50]| 43 & d6 \\
40 & e7 & cc & e0 & e6 & d1 & 9a & 1a & b3 & 28 & 1c & 7c & 0c & b9 & |[fill=matchColorFantomas!50]| c0 & 71 & 21 \\
50 & cb & 11 & 9e & |[fill=matchColorFantomas!50]| e3 & 48 & cd & e9 & 57 & f5 & 63 & 36 & 1b & b8 & bf & 9d & a7 \\
60 & 61 & d7 & f3 & |[fill=matchColorFantomas!50]| a9 & 12 & fd & |[fill=matchColorFantomas!50]| c1 & b7 & 8e & |[fill=matchColorFantomas!50]| a6 & 6b & 66 & |[fill=matchColorFantomas!50]| 72 & 64 & 85 & d5 \\
70 & |[fill=matchColorFantomas!50]| 4b & 7e & 67 & 3c & 65 & 17 & ba & 4a & |[fill=matchColorFantomas!50]| 97 & 29 & 83 & 6a & ae & f0 & e4 & 2e \\
80 & 77 & 74 & e8 & 2a & |[fill=matchColorFantomas!50]| ac & 95 & 3a & a2 & 3d & fa & 50 & 58 & ea & 9f & 93 & 33 \\
90 & b5 & 5c & 06 & |[fill=matchColorFantomas!50]| 51 & a3 & |[fill=matchColorFantomas!50]| 76 & 7a & 80 & bd & 16 & 39 & 0a & 03 & 73 & d0 & |[fill=matchColorFantomas!50]| 05 \\
a0 & f9 & b0 & 55 & 2d & b2 & 49 & f7 & |[fill=matchColorFantomas!50]| 19 & c6 & 45 & |[fill=matchColorFantomas!50]| d2 & d8 & 5d & f2 & 87 & ed \\
b0 & da & eb & 91 & ca & 3b & 47 & cf & |[fill=matchColorFantomas!50]| fb & c7 & dc & f6 & a4 & df & fe & b1 & |[fill=matchColorFantomas!50]| 09 \\
c0 & 0f & 0d & 2b & 26 & 14 & ff & 4d & |[fill=matchColorFantomas!50]| bc & |[fill=matchColorFantomas!50]| 02 & 81 & b4 & be & 15 & c5 & d4 & 27 \\
d0 & |[fill=matchColorFantomas!50]| 88 & 04 & 82 & c8 & 46 & e5 & 24 & c2 & 9b & 7d & 8d & d9 & 38 & 6d & ef & |[fill=matchColorFantomas!50]| a1 \\
e0 & dd & |[fill=matchColorFantomas!50]| 69 & |[fill=matchColorFantomas!50]| 5a & 54 & 9c & 53 & 25 & 20 & 5b & db & 37 & 5e & ab & 56 & 0b & 4c \\
f0 & 13 & 3e & 8a & d3 & ad & 31 & 08 & 96 & a5 & 18 & b6 & e2 & aa & 92 & c9 & bb \\
 	};
\end{tikzpicture}
\caption{An optimized Fantomas S-box using Algorithm \ref{algo}}
\label{fig:S_Fantomas_opt}
\end{figure}

\begin{figure}
\centering

\begin{tikzpicture}[scale=0.7, every node/.style={anchor=base,text depth=.4ex,text height=1ex,text width=2ex, scale=0.7}]

\matrix (A) [matrix of nodes, nodes={draw}, font=\ttfamily, row 1/.style={nodes={fill=gray!50}}, column 1/.style={nodes={fill=gray!50}}]
	{	& 00 & 01 & 02 & 03 & 04 & 05 & 06 & 07 & 08 & 09 & 0a & 0b  & 0c & 0d & 0e & 0f \\
		00 & |[fill=matchColorSkipjack!50]| a3 & |[fill=matchColorSkipjack!50]| d7 & |[fill=matchColorSkipjack!50]| 09 & |[fill=matchColorSkipjack!50]| 83 & |[fill=matchColorSkipjack!50]| f8 & |[fill=matchColorSkipjack!50]| 48 & |[fill=matchColorSkipjack!50]| f6 & |[fill=matchColorSkipjack!50]| f4 & |[fill=matchColorSkipjack!50]| b3 & |[fill=matchColorSkipjack!50]| 21 & |[fill=matchColorSkipjack!50]| 15 & |[fill=matchColorSkipjack!50]| 78 & |[fill=matchColorSkipjack!50]| 99 & |[fill=matchColorSkipjack!50]| b1 & |[fill=matchColorSkipjack!50]| af & |[fill=matchColorSkipjack!50]| f9 \\
10 & 5a & 6b & 69 & 0a & 0f & 27 & |[fill=matchColorSkipjack!50]| ca & 2f & |[fill=matchColorSkipjack!50]| 52 & |[fill=matchColorSkipjack!50]| 95 & c3 & 0d & |[fill=matchColorSkipjack!50]| 4e & a0 & c5 & 2c \\
20 & 8a & 12 & 38 & 6e & bb & d9 & c8 & e2 & cd & 02 & 7e & 5f & f1 & 87 & 19 & 8f \\
30 & |[fill=matchColorSkipjack!50]| 96 & fe & 2d & de & b2 & 6f & b6 & ac & 0c & |[fill=matchColorSkipjack!50]| ae & |[fill=matchColorSkipjack!50]| e5 & 7c & |[fill=matchColorSkipjack!50]| f7 & 43 & aa & 2a \\
40 & b9 & f3 & |[fill=matchColorSkipjack!50]| 7b & 1e & eb & 9a & cf & 73 & |[fill=matchColorSkipjack!50]| ee & 61 & |[fill=matchColorSkipjack!50]| 1a & a9 & 50 & 9b & ff & |[fill=matchColorSkipjack!50]| b8 \\
50 & 76 & 39 & 92 & 7f & 3a & 8c & |[fill=matchColorSkipjack!50]| bf & 14 & 60 & |[fill=matchColorSkipjack!50]| 58 & |[fill=matchColorSkipjack!50]| 80 & e1 & |[fill=matchColorSkipjack!50]| 66 & |[fill=matchColorSkipjack!50]| 0b & 86 & |[fill=matchColorSkipjack!50]| 90 \\
60 & 1f & 91 & 62 & ed & |[fill=matchColorSkipjack!50]| 33 & a6 & |[fill=matchColorSkipjack!50]| 65 & e0 & 67 & d4 & 82 & d6 & |[fill=matchColorSkipjack!50]| 6d & |[fill=matchColorSkipjack!50]| 98 & e6 & 74 \\
70 & e8 & 44 & 93 & |[fill=matchColorSkipjack!50]| c2 & |[fill=matchColorSkipjack!50]| b0 & fc & 9d & 6a & 81 & fa & 56 & 42 & 4d & 05 & |[fill=matchColorSkipjack!50]| 4a & 5c \\
80 & d8 & 5d & df & a4 & |[fill=matchColorSkipjack!50]| 49 & 1d & 9e & 16 & 4c & |[fill=matchColorSkipjack!50]| d2 & be & 00 & ba & c6 & 47 & 53 \\
90 & 84 & c0 & 55 & 3f & 1c & c7 & d5 & a2 & 88 & 34 & |[fill=matchColorSkipjack!50]| dc & |[fill=matchColorSkipjack!50]| c9 & 7d & 3c & |[fill=matchColorSkipjack!50]| 31 & 20 \\
a0 & d3 & 2e & e9 & 28 & 9c & 8e & 23 & ce & dd & 94 & 85 & a5 & 22 & 79 & a8 & |[fill=matchColorSkipjack!50]| 40 \\
b0 & 30 & |[fill=matchColorSkipjack!50]| 4b & e4 & 1b & |[fill=matchColorSkipjack!50]| d1 & 89 & a7 & |[fill=matchColorSkipjack!50]| 3b & 11 & c1 & fd & 36 & e7 & cc & |[fill=matchColorSkipjack!50]| 5b & 64 \\
c0 & 9f & |[fill=matchColorSkipjack!50]| 04 & a1 & 51 & 97 & 13 & 26 & |[fill=matchColorSkipjack!50]| f0 & |[fill=matchColorSkipjack!50]| 29 & db & cb & 7a & 75 & 8b & 77 & d0 \\
d0 & 3d & |[fill=matchColorSkipjack!50]| ef & |[fill=matchColorSkipjack!50]| bc & 70 & 71 & 63 & |[fill=matchColorSkipjack!50]| 37 & 2b & |[fill=matchColorSkipjack!50]| ec & 41 & da & 24 & ad & 8d & |[fill=matchColorSkipjack!50]| 10 & 18 \\
e0 & 01 & b5 & 54 & 07 & 35 & |[fill=matchColorSkipjack!50]| 4f & b7 & c4 & |[fill=matchColorSkipjack!50]| 32 & 17 & b4 & fb & 72 & 06 & ab & 0e \\
f0 & |[fill=matchColorSkipjack!50]| 5e & |[fill=matchColorSkipjack!50]| 6c & 68 & f2 & |[fill=matchColorSkipjack!50]| 57 & |[fill=matchColorSkipjack!50]| 25 & f5 & |[fill=matchColorSkipjack!50]| e3 & |[fill=matchColorSkipjack!50]| bd & 08 & 3e & 03 & 45 & |[fill=matchColorSkipjack!50]| 59 & ea & |[fill=matchColorSkipjack!50]| 46 \\
 	};
\end{tikzpicture}
\caption{An optimized Skipjack S-box using Algorithm \ref{algo}}
\label{fig:S_Skipjack_opt}
\end{figure}

\begin{figure}[]
\centering

\begin{tikzpicture}[scale=0.7, every node/.style={anchor=base,text depth=.4ex,text height=1ex,text width=2ex, scale=0.7}]

\matrix (A) [matrix of nodes, nodes={draw}, font=\ttfamily, row 1/.style={nodes={fill=gray!50}}, column 1/.style={nodes={fill=gray!50}}]
	{	& 00 & 01 & 02 & 03 & 04 & 05 & 06 & 07 & 08 & 09 & 0a & 0b  & 0c & 0d & 0e & 0f \\
		00 & ab  &  f0  &  5e  &  3f  &  fa  &  e2  &  6f  &  8e  &  3c  &  36  &  30  &  db  &  29  &  73  &  da  &  45  \\
10 & 87  &  f9  &  60  &  3b  &  bf  &  a4  &  c7  &  0c  &  a9  &  c0  &  f3  &  cb  &  68  &  ff  &  ee  &  a6  \\
20 & 90  &  57  &  f2  &  77  &  ef  &  c2  &  78  &  b7  &  94  &  32  &  e6  &  4d  &  53  &  6d  &  26  &  98  \\
30 & c1  &  2c  &  2a  &  9a  &  12  &  2b  &  ea  &  e8  &  17  &  7c  &  5c  &  6e  &  50  &  d9  &  f6  &  88  \\
40 & 83  &  69  &  5a  &  67  &  af  &  b9  &  1a  &  b8  &  8a  &  d4  &  b4  &  a0  &  cc  &  e1  &  24  &  c6  \\
50 & be  &  1f  &  a1  &  51  &  9f  &  64  &  4e  &  4f  &  2f  &  85  &  6b  &  76  &  86  &  35  &  4b  &  ed  \\
60 & 81  &  84  &  39  &  13  &  62  &  c3  &  9e  &  dc  &  d0  &  66  &  5f  &  44  &  de  &  1c  &  bd  &  34  \\
70 & 1d  &  1e  &  2d  &  6c  &  a2  &  46  &  97  &  c5  &  37  &  61  &  a3  &  56  &  fe  &  f7  &  d5  &  38  \\
80 & ce  &  05  &  09  &  18  &  aa  &  fc  &  91  &  28  &  9b  &  10  &  e9  &  0b  &  71  &  dd  &  e7  &  23  \\
90 & 7f  &  72  &  59  &  6a  &  43  &  fd  &  d1  &  e4  &  f8  &  0d  &  55  &  74  &  c8  &  f5  &  27  &  65  \\
a0 & 93  &  c4  &  19  &  49  &  00  &  20  &  3d  &  2e  &  a8  &  d3  &  01  &  7d  &  25  &  0e  &  f4  &  33  \\
b0 & 02  &  04  &  0a  &  14  &  16  &  ae  &  31  &  11  &  cf  &  79  &  8f  &  d8  &  8b  &  d7  &  ca  &  b3  \\
c0 & bb  &  3e  &  0f  &  92  &  df  &  40  &  4c  &  cd  &  ac  &  22  &  5b  &  a5  &  bc  &  f1  &  75  &  89  \\
d0 & 96  &  b1  &  e3  &  d2  &  7a  &  1b  &  70  &  58  &  03  &  47  &  80  &  9c  &  06  &  ba  &  c9  &  54  \\
e0 & ad  &  41  &  99  &  48  &  7e  &  3a  &  95  &  e0  &  ec  &  07  &  63  &  7b  &  b2  &  21  &  b0  &  4a  \\
f0 & 8d  &  d6  &  15  &  fb  &  9d  &  5d  &  8c  &  42  &  08  &  b6  &  eb  &  a7  &  b5  &  e5  &  52  &  82  \\
 	};
\end{tikzpicture}
\caption{An optimized S-box $S_c(8,8)$  using Algorithm \ref{algo}}
\label{fig:Sc_raw}
\end{figure}

\begin{table}
\centering
\caption{Nonlinearities of $S_c$ by coordinates}
\label{tab:proposed8}       
\scalebox{0.9}{
\begin{tabular}{rllllllll}
 & \multicolumn{8}{c}{} \\
 Method & $f_1$ & $f_2$ & $f_3$ & $f_4$ & $f_5$ & $f_6$ & $f_7$ & $f_8$ \\
\noalign{\smallskip}\hline\noalign{\smallskip}
 this work & 114 & 114 & 114 & 114 & 114 & 114 & 114 & 114 \\
\noalign{\smallskip}\hline
\end{tabular}}
\end{table} 

\begin{table}
\begin{center}
\caption{SAC, Coordinate-average and final nonlinearity of $S_c$}
\label{tab:proposedNL}       
\scalebox{1}{
\begin{tabular}{rllccc}
\hline
 Method & min & max & ACNV & SAC & NL \\
\noalign{\smallskip}\hline\noalign{\smallskip}
 this work & 114 & 114 & 114 & 0.5000000 & 96 \\
\noalign{\smallskip}\hline
\end{tabular}}
\end{center}
\end{table}

\section{S-box as multi-armed bandits}
\label{sec:bandits}

The space of bijective S-boxes is vast. For example, in the case of (8,8) S-boxes, we have a total of $256! \approx 2^{1684}$ different bijective S-boxes. Despite algorithm \ref{algo} efficiency in finding S-boxes with better ACNV, we had never reached an ACNV greater than 114. However, we have found out that the multi-armed bandit problem \cite{berry1985bandit}\cite{katehakis1987multi}\cite{li2016art}\cite{li2016collaborative} is closely related to the nonlinearity optimization problem.

Each bijective S-box $S(n,n)$ can be represented as a collection of $n$ bandits, such that each bandit uniquely corresponds to some of the $n$ coordinates of $S$. The arms of each bandit could be associated with the operation of applying a single transposition in $S$, while the profit of our action could be measured with the fitness function presented in Algorithm \ref{algo}. 

Associating each one of the possible $n \choose {2}$ transposition of elements of $S$ DLUT to some distinct arm in each bandit is a trivial and non-working model - at the end, the bandits would be indistinguishable. Having this in mind, the following model is constructed:

\begin{itemize}
\item{\textbf{Property I}: Since each bandit uniquely corresponds to some coordinate of $S$, each bandit arm is restricted to initiate a transposition of two bits inside a column of $S_{LUT}$ only (instead of a transposition of any two elements in $S_{DLUT}$).}
\item{\textbf{Property II}: To keep the bijective property of $S$, in case an arm of some bandit is activated, the set of all distinct ${2^n \choose 2}$ bit transpositions in a given coordinate of $S_{LUT}$ is restricted to a subset of transpositions with a size of $2^{n-1}$.}
\end{itemize}

The restriction introduced in \textbf{Property II} is motivated by the following observations:

\begin{enumerate}
\item{\textbf{Existence}: If $b_1b_2 \cdots b_i \cdots b_n$ is a row from $S_{LUT}$, flipping the bit $b_i$ will result in \textbf{some} other row $R = b_1b_2 \cdots \overline{b_i} \cdots b_n$ in $S_{LUT}$. Otherwise, if $R$ is not among the rows of $S_{LUT}$, $S$ is not surjective, therefore not bijective, which contradicts our initial choice of $S$.}
\item{\textbf{One-to-one Maping:} If $b_1b_2 \cdots b_i \cdots b_n$ is a row from $S_{LUT}$, flipping the bit $b_i$ will result in \textbf{only one} row $R = b_1b_2 \cdots \overline{b_i} \cdots b_n$ in $S_{LUT}$. Otherwise, if some other row $R^{'}$ of $S_{LUT}$ exists, s.t. $R \equiv R^{'}$, $S$ is not injective, therefore not bijective, which contradicts our initial choice of $S$.}
\item{\textbf{Search space:} The total number of distinct bit sequences of the form $b_1b_2 \cdots b_{i-1}b_{i+1} \cdots b_n$ is $2^{n-1}$.}
\end{enumerate}

Let's denote as a bandit $B_i$ the bandit, which corresponds to the $i$-th coordinate of $S$. Each bandit consists of $2^{n-1}$ distinct arms, s.t. each arm of $B_i$ corresponds to a distinct value of $b_1b_2 \cdots b_{i-1}b_{i+1} \cdots b_n$. Activating an arm of $B_i$ will result of interchanging two rows of $S_{LUT}$, which differ only in bit position $i$. 

\begin{figure*}
\centering
  \includegraphics[width=0.7\textwidth]{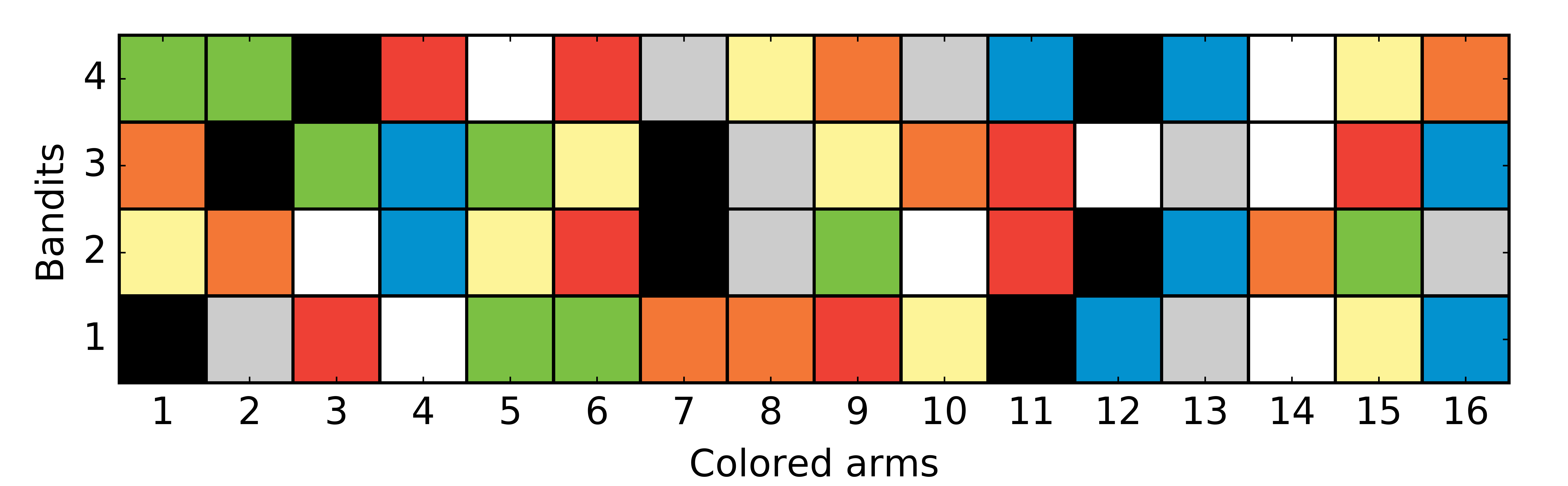}
\caption{An example of 8-armed 4-bandit problem transformation of the S-box $X$}
\label{fig:bandits}       
\end{figure*}

For example, let's consider an S-box X(4,4), with $X_{DLUT} = $ [15, 14, 9, 2, 11, 3, 12, 4, 1, 13, 7, 8, 6, 10, 5, 0]. $X$ is a bijective S-box with dimension $4$. Therefore, we can transform $X$ as an $8$-armed $4$-bandit problem. In Figure \ref{fig:bandits}, a visual interpretation of the bandits transformation of $X$ is shown. Each row corresponds to a distinct bandit, while each pair of cells inside a given row, sharing the same color, corresponds to an arm of the given bandit. The x-axis represents the indexes of elements of $X_{DLUT}$ (starting from 1).

As an illustration, if we activate the white arm of bandit 1, we interchange the elements of $X_{DLUT}$ with indexes 14 and 4, i.e. 10 and 2. Their respective binary representations (with zero-fill of 4) are \textbf{1010} and \textbf{0010} (they differ only in bit position 1).

The profit (if any) of activating a bandit $B_i$'s arm is measured by the same function $E$ presented in Algorithm \ref{algo}.

The transformation of the (n,n) bijective S-box ACNV optimization problem to the $2^{n-1}$-armed $n$-bandit problem allows us to focus on the optimization of the nonlinearity of single coordinates. Furthermore, by design, activating an arm of a given bandit doesn't affect the states of other bandits. Having this in mind, Algorithm \ref{algoBandit} is proposed.

\begin{algorithm}
\caption{}\label{algoBandit}
\begin{algorithmic}[1]
\State $s\gets R(n)$ \Comment{the function R(n) generates pseudo-random bijective S-box $S(n,n)$}
\State $\Omega \gets MODEL(s)$ \Comment{We transform the S-box $s$ to a $2^{n-1}$-armed $n$-bandit problem}
\Repeat
	\State $bandit \gets random(1,n,\Lambda)$ \Comment{We choose a random bandit from $\left[1,n \right]$, based on some profit-maximizing strategy $\Lambda$}
	\State $arm \gets random(1,2^{n-1})$ \Comment{We choose a random arm from $\left[1,2^{n-1} \right]$}
	\State $oldBandit \gets \Call{E}{bandit}$
	\State $\Call{Activate}{bandit, arm}$ 
	\If{$\Call{E}{bandit}  < \Call{E}{oldBandit}$}
		\State $\Omega \gets MODEL(s)$ \Comment{We update the model}
	\Else{}
		\State $\Call{Activate}{bandit, arm}$ \Comment{We resume the original state of the bandit}
	\EndIf	
\Until{STOP condition is reached}  \Comment{reaching $n{2^{n-1}}$ consequent unsuccessful attempts}
\end{algorithmic}
\end{algorithm}

\section{Results Part II}

Our implementation of Algorithm \ref{algoBandit} is based on a simple strategy $\Lambda$ - we always choose a bandit, which posses the lowest nonlinearity. In case there are several bandits sharing the lowest value of  nonlinearity, we choose one of them at random. 

We launched Algorithm \ref{algoBandit} as a stand-alone optimization routine, starting from pseudo-randomly generated S-boxes, and in almost all of the instances we reached S-boxes with an average coordinate nonlinearity value of 112. However, when we initiated Algorithm \ref{algoBandit} with S-boxes, which have been already optimized by Algorithm \ref{algo}, we have reached an average coordinate nonlinearity value of 114.5. An example of such S-box is given in Figure \ref{fig:Sc_best}. The corresponding nonlinearity by coordinates is given in Table \ref{tab:114.5}. 

\begin{figure}[]
\centering

\begin{tikzpicture}[scale=0.7, every node/.style={anchor=base,text depth=.4ex,text height=1ex,text width=2ex, scale=0.7}]

\matrix (A) [matrix of nodes, nodes={draw}, font=\ttfamily, row 1/.style={nodes={fill=gray!50}}, column 1/.style={nodes={fill=gray!50}}]
	{	  & 00 & 01 & 02 & 03 & 04 & 05 & 06 & 07 & 08 & 09 & 0a & 0b  & 0c & 0d & 0e & 0f \\
00 & a9 & 7b & bb & c9 & 0a & 55 & d1 & c1 & a3 & 1a & 24 & 26 & bf & 72 & f6 & 0d \\
10 & 4a & 73 & e2 & f1 & 3a & d3 & 35 & b2 & 64 & 93 & f5 & d7 & ff & dc & cb & 4e \\
20 & de & 7a & a2 & 98 & d9 & 87 & b3 & a5 & 28 & ba & a0 & 45 & 56 & 67 & 61 & 0c \\
30 & a7 & 33 & 53 & 2d & e7 & 58 & 7e & b6 & 37 & 71 & 1e & 10 & d0 & e0 & b7 & 49 \\
40 & 96 & 91 & 6e & b0 & f9 & 5b & fb & 13 & 8a & db & ad & a1 & 8c & 39 & 22 & ee \\
50 & 89 & 4f & 50 & da & 07 & 75 & 65 & bd & 9f & 18 & cd & 17 & 41 & be & 2f & 40 \\
60 & ca & 05 & ae & 32 & 94 & f7 & a8 & b1 & aa & f8 & e9 & e4 & 82 & 54 & 01 & 69 \\
70 & 27 & 81 & 5c & 84 & 7f & b4 & 29 & d2 & fc & 0e & a4 & 36 & 90 & 2e & 15 & 00 \\
80 & 4c & 51 & 25 & 11 & 16 & f3 & 3d & 8d & 9c & 6c & 95 & ef & 76 & cc & 8b & dd \\
90 & 2b & f4 & ce & 43 & 62 & d4 & 74 & fe & 92 & c2 & 7c & 80 & 2a & 21 & 68 & bc \\
a0 & c0 & 23 & af & e3 & 78 & 6f & e1 & eb & 03 & 38 & 09 & 42 & d6 & ed & ec & 02 \\
b0 & 1d & fa & 5e & b9 & c8 & c7 & 46 & 14 & e6 & 99 & 9d & 04 & c4 & d8 & 3f & 9b \\
c0 & e5 & 4d & 31 & 63 & 79 & 3c & d5 & f0 & 47 & 57 & 4b & c6 & f2 & 2c & 70 & b5 \\
d0 & cf & 9e & 0b & 3e & 1f & 5a & a6 & 6a & 6b & 12 & 1b & 77 & 5f & 48 & ac & 3b \\
e0 & 44 & 34 & 5d & ea & 20 & 85 & 8f & 30 & 9a & ab & 1c & c3 & 59 & 8e & fd & 08 \\
f0 & b8 & e8 & 06 & 6d & 66 & 7d & df & 60 & 52 & 83 & 88 & 19 & 0f & 97 & c5 & 86 \\
 	};
\end{tikzpicture}
\caption{An optimized, by applying the composition of Algorithms \ref{algo} and \ref{algoBandit}, (8,8) S-box}
\label{fig:Sc_best}
\end{figure}

\begin{table}
\centering
\caption{Nonlinearities of the S-box coordinates given in Figure \ref{fig:Sc_best}}
\label{tab:114.5}       
\scalebox{0.9}{
\begin{tabular}{rllllllll}
 & \multicolumn{8}{c}{} \\
 Method & $f_1$ & $f_2$ & $f_3$ & $f_4$ & $f_5$ & $f_6$ & $f_7$ & $f_8$ \\
\noalign{\smallskip}\hline\noalign{\smallskip}
 this work & 116 & 114 & 116 & 114 & 114 & 114 & 114 & 114 \\
\noalign{\smallskip}\hline
\end{tabular}}
\end{table} 

In Table \ref{tab:summaryResults}, an extended S-box comparison between the state-of-the-art methods is given. The entries are sorted, in increasing order, by ACNV (the last column). 

\begin{table}
\centering
\caption{A comparison of S-boxes, yielded by various methods to be found in the literature, with those S-boxes, reached by the algorithms presented in this work}
\label{tab:summaryResults}       
\rowcolors{2}{gray!25}{white}
\begin{tabular}{lllll}

 & \multicolumn{3}{c}{} \\
 Method & Min NL & Max NL & ACNV \\
\noalign{\smallskip}\hline\noalign{\smallskip}
 \cite{khan2015novel}\cite{khan2016construction} & 84 & 106 & 100.0 \\
\cite{jamal2016watermarking} & 98 & 108 & 102.3 \\
\cite{khan2018novel} & 96 & 106 & 102.5 \\
\cite{chen2007extended} & 100 & 106 & 103.0 \\
\cite{khan2012novel} & 96 & 106 & 103.0 \\
\cite{khan2013efficient} & 98 & 108 & 103.0 \\
\cite{jakimoski2001chaos} & 98 & 108 & 103.2 \\
\cite{ozkaynak2010method} & 100 & 106 & 103.2 \\
\cite{tang2005novel} & 99 & 106 & 103.3 \\
\cite{asim2008efficient} & 96 & 108 & 103.5 \\
\cite{tang2005method} & 101 & 108 & 103.8 \\
\cite{ozkaynak2013designing} & 101 & 106 & 103.8 \\
\cite{chen2008novel} & 102 & 106 & 104.0 \\
\cite{khan2013efficient2} & 98 & 108 & 104.0 \\
\cite{khan2014construction} & 100 & 106 & 104.0 \\
\cite{liu2014chaos} & 102 & 106 & 104.0 \\
\cite{khan2016new} & 98 & 108 & 104.0 \\
\cite{liu2018novel} & 102 & 108 & 104.5 \\
\cite{hussain2013group}\cite{ozkaynak2017new} & 100 & 108 & 104.7 \\
\cite{hussain2012novel}\cite{ozkaynak2019analysis} & 102 & 108 & 104.7 \\
\cite{khan2015efficient} & 100 & 108 & 104.75 \\
\cite{hussain2012construction} & 100 & 107 & 104.8 \\
\cite{ozkaynak2017biometric} & 104 & 106 & 105.0 \\
\cite{hussain2013novel} & 102 & 108 & 105.2 \\
\cite{belazi2017efficient} & 102 & 108 & 105.3 \\
\cite{hussain2013projective} & 100 & 110 & 105.5 \\
\cite{khan2014novel} & 98 & 110 & 105.5 \\
\cite{belazi2017simple} & 102 & 110 & 105.5 \\
\cite{liu2015designing} & 104 & 108 & 105.7 \\
\cite{ul2017designing} & 102 & 108 & 106.0 \\
\cite{ccavucsouglu2018novel}\cite{wang2018chaotic}\cite{wang2019s}a & 104 & 110 & 106.0 \\
\cite{wang2019s}c & 106 & 108 & 106.0 \\
\cite{ccavucsouglu2017novel} & 104 & 110 & 106.2 \\
\cite{farah2017novel} & 104 & 110 & 106.5 \\
\cite{lambic2018s} & 106 & 108 & 106.5 \\
\cite{ozkaynak2019construction}\cite{lambic2017novel} & 106 & 108 & 106.7 \\
\cite{ye2018chaotic} & 104 & 108 & 106.7 \\
\cite{ahmad2015novel} & 106 & 110 & 107.0 \\
\cite{wang2019s}b & 104 & 108 & 107.0 \\
\cite{ahmed2019novel} & 106 & 108 & 107.5 \\
\cite{yi2019novel} & 106 & 110 & 107.75 \\
\cite{wang2012novel} & 108 & 108 & 108.0 \\
\cite{zhang2014chaotic} & 104 & 110 & 108.0 \\
\cite{al2018new} & 106 & 110 & 108.5 \\
\cite{lambic2014novel} & 108 & 112 & 109.0 \\
\cite{hussain2013efficient}\cite{hussain2013efficient2}\cite{belazi2017chaos} & 112 & 112 & 112.0 \\
Alg.2 (this work) & 112 & 112 & 112.0 \\
Alg.1 (this work)& 114 & 114 & 114.0 \\
Alg.1 $\bigcup$ Alg.2 (this work) & 114 & 116 & 114.5 \\
\noalign{\smallskip}\hline
\end{tabular} 
\end{table} 

\section{Conclusion and future work}
\label{sec:6}
CF-based S-box construction is a relatively new and interesting technique, which interconnects the tools provided by various academic disciplines with the problem of finding secure cryptographic primitives. 

In this paper, we analyzed the actual linear cryptanalysis resistance of CF-based S-boxes, which differs from the average nonlinearity value announced by a great number of papers. Integrating such S-boxes in a cryptosystem should be done with a considerable caution. For example, if we interchange the Rijndael S-box in AES \cite{AES} with some CF-based S-box with higher ACNV, but lower overall nonlinearity, the resulting modified block cipher will be significantly weaker in terms of resistance to linear cryptanalysis. Furthermore, we show that exploiting chaos structures, in the context of nonlinearity optimization problem, is arguable. Thus, the benefits of using chaos structures in the design of S-boxes is unclear and yet to be determined. However, as stated in \cite{accikkapi2019side}, the chaos-based designs may be an alternative to application attacks, such as side-channel analysis.

Nevertheless, from designer perspective, if the overall nonlinearity value of an S-box $S$ is negligible compared to the average nonlinearity value of all coordinates of $S$, two novel S-box constructions are suggested. 

While Algorithm \ref{algo} yields better results than Algorithm \ref{algoBandit}, the latest could be used as an Algorithm \ref{algo} extension, to further improve the parameters of the resulting S-box. The methods presented in this paper significantly outperform all other state-of-the-art methods for designing S-boxes with high ACNV. 

The linkage of the $n$-armed bandit problem to the problem of finding such S-boxes, opens an interesting area of future research - the investigation of how other state-of-the-art methods, such as the concept of fuzzy graphs \cite{rosenfeld1975fuzzy}\cite{hao2016mining}, the stochastic optimization techniques \cite{ermoliev1988numerical}\cite{thathachar2011networks}\cite{narasimhan2016stochastic}, or the exploration-exploitation algorithms \cite{alba2005exploration}\cite{macready1998bandit}\cite{li2015data}, could be exploited to further maximize the ACNV of a given S-box.

An interesting open question to be answered is to what extend the ACNV value of an (8,8) bijective S-box could be optimized? As summarized in \cite{picek2016maximal}, the maximal nonlinearity value achieved in balanced boolean functions with 8 variables is 116. Therefore, if an ACNV for an (8,8) bijective S-box greater than 116.0 is ever found, at least one of its eight components will posses nonlinearity value 118, which will finally give an answer to the long-standing problem of the maximum possible nonlinearity value of an eight variable balanced boolean functions. Furthermore, as shown in \cite{sarkar2000nonlinearity}, the upper bound for eight variable balanced boolean functions is less than 120. Thus, the maximum theoretical possible ACNV of (8,8) bijective S-boxes is less or equal to 118.0, but most probably, considering the academic skepticism that eight variable balanced boolean functions with nonlinearity value 118 really exist, less or equal to 116.0.

\bibliographystyle{IEEEtran}
\bibliography{refs}

\begin{IEEEbiography}[{\includegraphics[width=1in,height=1.25in,clip,keepaspectratio]{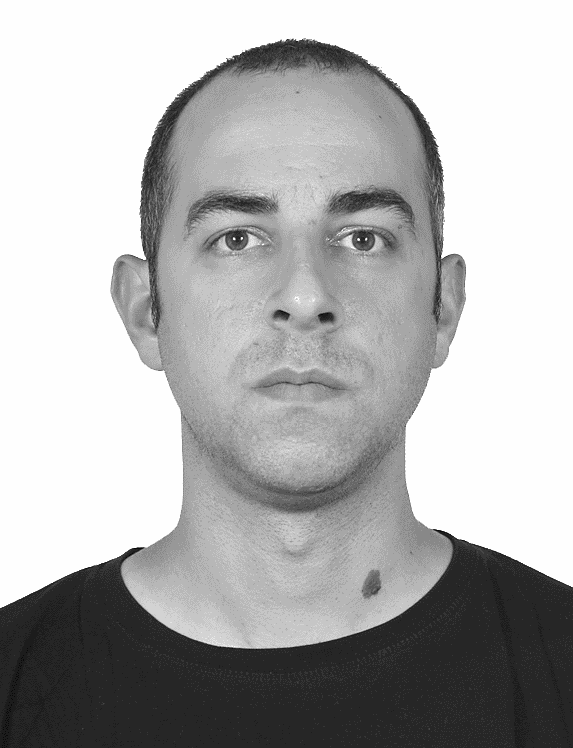}}]{Miroslav M. Dimitrov} was born in Yambol, Bulgaria in 1985. He received the B.S. degree in Informatics and M.S. degree in Information Security, both from the Faculty of Mathematics and Informatics, Sofia University. He is currently pursuing the Ph.D. degree in Informatics at Bulgarian Academy of Sciences. His research interests include cryptology, algorithms, and sequences.
\end{IEEEbiography}

\EOD

\end{document}